\documentstyle[psfig]{mn}
\begin{document}
\def\simlt{\mathrel{\rlap{\lower 3pt\hbox{$\sim$}}
        \raise 2.0pt\hbox{$<$}}}
\def\simgt{\mathrel{\rlap{\lower 3pt\hbox{$\sim$}}
        \raise 2.0pt\hbox{$>$}}}

\title[The Redshift Evolution of Clustering in the HDF]
{The Redshift Evolution of Clustering in the HDF}
\author[M.Magliocchetti, S.J.Maddox]
{M.Magliocchetti$^{1}$, S.J.Maddox$^{1,2}$ \\
$^1$Institute of Astronomy, Madingley Road, Cambridge CB3 0HA\\
$^2$School of Physics and Astronomy, University of Nottingham,
Nottingham, NG7 2RD, UK.}

\maketitle
\begin{abstract}

We present a correlation function analysis for the catalogue of
photometric redshifts obtained from the Hubble Deep Field image by
Fernandez-Soto et al., 1998.
By dividing the catalogue into redshift bins of width $\Delta z=0.4$
we measured the angular correlation function $w(\theta)$ as a function
of redshift up to $z\sim 4.8$.
From these measurements we derive the trend of the correlation length $r_0$.  
We find that $r_0(z)$ is roughly constant with look-back time up to
$z \simeq 2$, and then increases to higher values at $z\simgt 2.4$. 
We estimate the values of $r_0$, assuming 
$\xi(r,z)=(r/r_0(z))^{-\gamma}$, $\gamma=1.8$ and different
geometries. For $\Omega_0=1$ we
find $r_0(z=3)\simeq 7.00\pm 4.87\; h^{-1}$ Mpc, 
in good agreement with the values obtained from
analysis of the Lyman Break Galaxies.

\end{abstract}
\section{INTRODUCTION}

The evolution of galaxy clustering provides vital clues to the
formation of galaxies and large-scale structure. The amplitude of galaxy
clustering is determined by the combination of the evolution of the
underlying mass fluctuations, and the bias relating the galaxy
overdensities to mass.  Observationally the amplitude of the galaxy
correlation function has been measured from redshift surveys extending
up to redshifts $z\sim 1$ (e.g. CFRS, Le Fevre et al., 1995; CNOC2,
Carlberg et al. 1998). 
The discovery of Lyman Break Galaxies (Steidel et al., 1996) has
allowed the scientific community to push this limit even further up to
$z\simeq 3$.  However there is still a gap between the measurements
obtained for $z\simlt 1$ and those provided by the analysis of the
Lyman Break Galaxies at redshifts $z=3$.

The aim of this letter is to ``fill in'' this gap by presenting
measurements of the correlation function obtained from the catalogue
of photometric redshifts derived from the Hubble Deep Field by
Fernandez-Soto et al., 1998.  Using the photometric redshifts we 
divide the catalogue into subsamples in  redshift and measure the
angular correlation function $w(\theta)$ and 
its amplitude as a function of $z$ up to $z\simeq 4.8$.  At low redshift
our results agree with the results of Connolly et al. (1998), and at
higher redshifts they are consistent with Lyman Break Galaxies.

The outline of the paper is as follows: section 2 gives a description
of the catalogue adopted for our analysis while section 3 presents
the results for the angular correlation function. In section 4 we
derive the meaningful spatial quantities as a function of redshift,
while section 5 summarises our conclusions.

\section{THE DATA}
The Hubble Deep Field (HDF) image (Williams et al., 1996) covers an
L-shaped area roughly  $ 3' \times 3' $, with a total area $\sim 4$arcmin$^2$ 
 and provides us with the
deepest view of the Universe obtained so far. The image was obtained
by the Hubble Space Telescope using the Wide Field Planetary Camera 2
(WFPC2) over a period of 10 consecutive days. The same image was
acquired by using the broad-band filters F300W, F450W, F60W and F814W
in order to allow for the possibility of assigning photometric
redshifts to the objects in the field.

Since then various techniques for getting photometric redshifts have
been applied to the HDF (see e.g. Connolly et al., 1998). The catalogue
we will use for this work has been derived by Fernandez-Soto et al.,
1998 by incorporating in their former technique (Lanzetta et al.,
1996) the infrared images of the HDF acquired in the
J(1.2$\mu$m), H(1.65$\mu$m) and K(2.2$\mu$m) broad-band filters
(Dickinson et al., 1998).
\begin{figure}
\vspace{8cm}  
\includegraphics{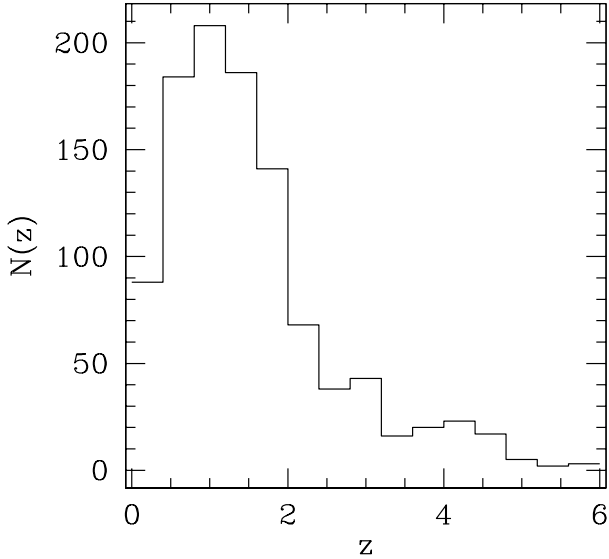} 
\caption{ Redshift distribution of the objects in the HDF as obtained
  by Fernandez-Soto et al., 1998
\label{fig:N_z} }
\end{figure}
Their final catalogue includes 1067 objects, some of them at very high
redshifts ($z_{max}\sim 6$), as seen in the redshift distribution shown
in Figure \ref{fig:N_z}.  
The edges of the Wide Field Camera images are of poorer quality than
the bulk of the images, and Fernandez-Soto et al. use a magnitude
limit of AB(8140)=28 for the inner part of the HDF, while the outer
part of the image includes only objects with AB(8140)$<$26.
Using the sensitivity map given in Fernandez-Soto et al., 1998 we
rejected all those galaxies belonging to the shallower region; our
final version of the catalogue, complete down to the magnitude
AB(8140)=28, includes 946 objects. 
Figure~1 shows the distribution of galaxies in the catalogue split
into a series of narrow redshift intervals. 

\section{THE ANGULAR CORRELATION FUNCTION}

\begin{figure}
\vspace{9cm}  
\includegraphics{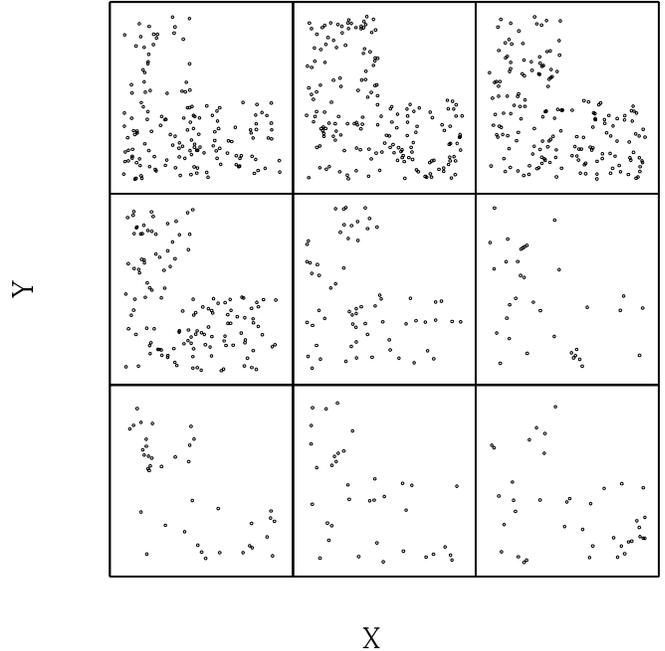} 
\caption{Spatial distribution of objects in the HDF with AB(8140)$<$28
in different redshifts intervals: $0.4-0.8$,
$0.8-1.2$, $1.2-1.6$, $1.6-2.0$,
$2.0-2.4$, $2.4-2.8$, $2.8-3.2$,
$3.2-4.0$ and $4.0-4.8$.  The resdhift binning is
as in Figure \ref{fig:w}}
\label{fig:dist}
\end{figure}

Correlation-function analysis has become the standard way to quantify
the clustering of different populations of astronomical sources.
Ideally we would like to measure the spatial correlation function, but
photometric redshifts are not precise enough to enable a direct
measurement: the typical error between the photometric estimates and
the spectroscopic measurements of $z$ is $\Delta z_{rms}\simeq
0.1\cdot (1+z)$ ($\Delta z=z_{sp}-z_{phot}$) (see Fernandez-Soto et
al., 1998), so that for instance $\Delta z_{rms}\sim 0.4$ at $z=3$,
corresponding to several hundred Mpc.
Nevertheless the estimated redshifts can be used to select subsamples
of galaxies at different redshifts, and so we can obtain estimates of
the spatial clustering as a function of redshift via the angular
correlation function.
As pointed out by Connolly at al (1998), the angular correlation
function from a redshift limited sample has much higher signal-to-noise
than a comparable magnitude limited sample. 

The angular two-point correlation function
$w(\theta)$ gives the excess probability, with respect to a
random Poisson distribution, of finding two sources in the solid
angles $\delta\Omega_1$ $\delta\Omega_2$ separated by an angle
$\theta$, and it is defined as
\begin{eqnarray} 
\delta P=n^2\delta\Omega_1\delta\Omega_2\left[1+w(\theta)\right]
\label{eqn:wthetadef} 
\end{eqnarray}
where $n$ is the mean number density of objects in the catalogue under
consideration.
One of the major limitations on the study of Large-Scale Structure
with the HDF is its small field of view; for $\Omega_0=1$, 220 arcsecs
correspond to 0.9$h^{-1}$ Mpc at $z=1$, (see Connolly et al., 1998), so
isolating narrow redshift intervals will select galaxies in a 
small volume, and so lead to very large errors in the clustering
analysis (e.g. a single cluster of galaxies could dominate the
signal). We therefore
decided to divide the sample into bins of width $\Delta z=0.4$ and
consider for our analysis all the objects in the ``clean'' 
catalogue (see section 2) up to redshifts $z=4.8$. 
This redshift bin width corresponds to between one and two times
the expected {\em rms} error in the redshifts. 
Given the small number of galaxies in the catalogue it might be
thought that the signal-to-noise could be improved by using broader
bins to increase the number of objects per bin and so reduce the
Poisson noise. However,
increasing the redshift range in each bin would lead to a noticeable
  reduction of the clustering signal due to the ``washing out'' of
  real structures caused by projection effects. As we will see later, 
  cosmic variance is not a problem in our analysis. In
  fact, the presence of an overdense (underdense) region in the
  sample would result in a sudden rise (fall) of the clustering amplitude
  only in one particular redshift bin. The smooth trend of the
  correlation length $r_0$ as a function of $z$ (see Figure 4) suggests 
  this is not the case in our analysis.

\begin{figure*}
\vspace{15cm}  
\includegraphics{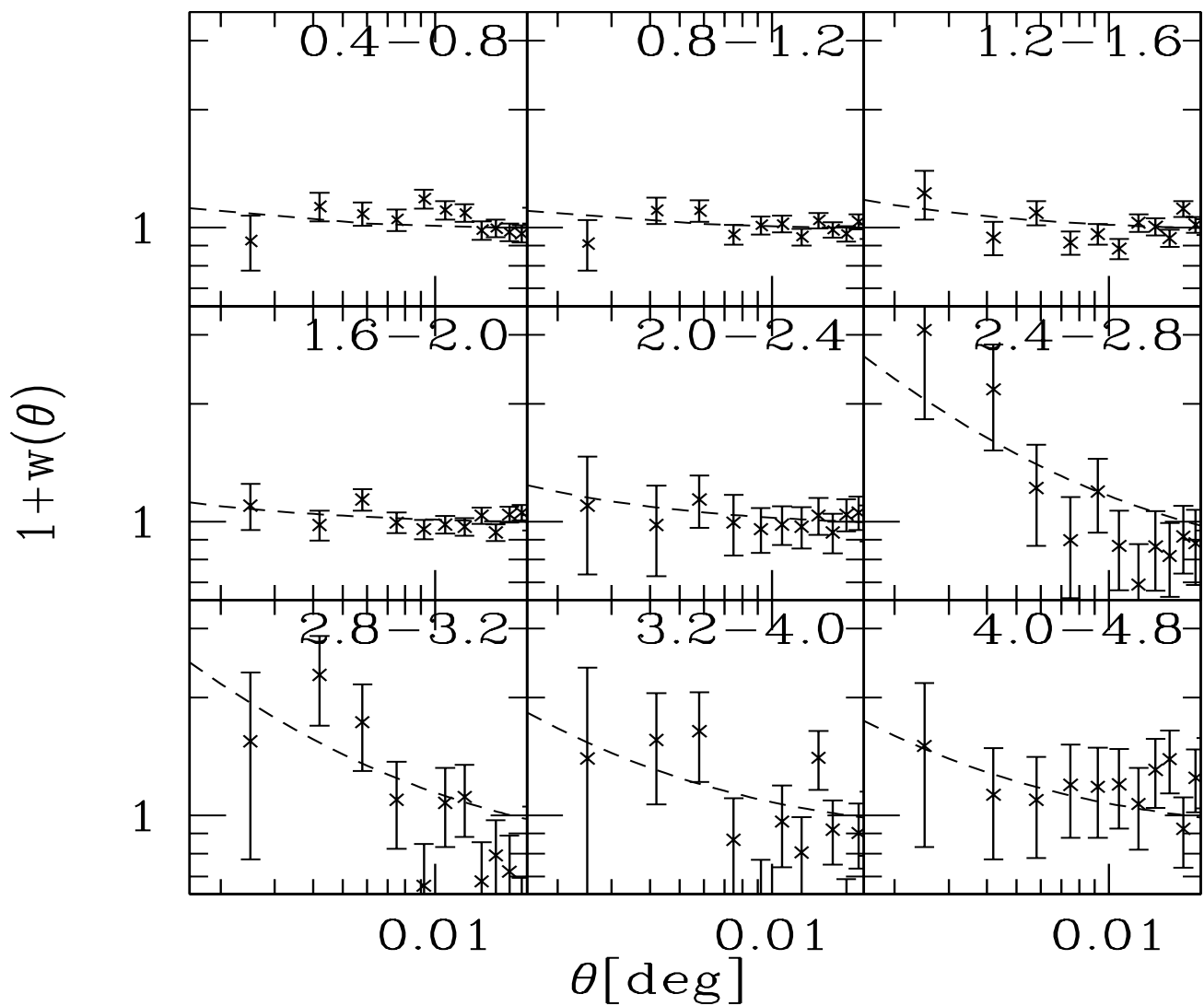}
\caption{The angular correlation function $w$ for galaxies within the HDF
with AB(8140)$<$28 at different redshifts intervals. The dashed lines
show the best fits to the data.
\label{fig:w} }
\end{figure*}

We then generated random catalogues containing 10000 galaxies, with
positions of the random objects lying within the area defined by the
geometry of the photometric data, for each of the subsamples
associated to a particular redshift bin and then counted the number of
distinct data-data pairs ($DD$), data-random pairs ($DR)$, and distinct
random-random pairs ($RR$) as a function of angular separation.  We
then calculate $w$ using  the estimator (Hamilton, 1993) 
\begin{eqnarray}
w = &\frac{4DD \ RR}{DR^2}  &-1 
\label{eqn:wtheta_ests} 
\end{eqnarray}
in the angular scales $9\le \theta \le 180$ arcsecs. 
We also used the estimators suggested by Peebles (1980) and Landy \& Szalay
(1993), and found virtually identical results. 
In Figure~\ref{fig:w} 
we show the results for $w$ for different redshift bins; 
the error bars show Poisson
estimates for the points. Since the distribution is clustered, these estimates
only provide a lower limit to the uncertainties. 
Nevertheless it can be shown that, over the range
of scales considered for the calculation of $w(\theta)$, Poisson
errors are comparable to those obtained from bootstrap resampling
(Villumsen et al., 1997, see also Connolly et al., 1998).

Note that in our analysis we did not include the results for $0\le
z\le 0.4$; this is due to the fact that the effect of excluding bright
(nearby) galaxies in the construction of the HDF sample results 
in a spurious reduction of the clustering amplitude in that redshift
range (Connolly et al., 1998).

As shown in Figure~\ref{fig:w} the angular correlation function is
significantly positive at small scales for all the redshift bins
considered in our analysis.  
The clustering amplitude is roughly constant within the errors up to
$z\sim 2.4$; above this value the amplitude increases, and remains
high for all $z >2.4$ samples.

The apparent clustering in the higher redshift samples may be partly
affected by the problem of defining single galaxies from the irregular 
morphology of some of the objects. However, this is a problem only for
scales smaller than $3''$, and we can see from visual
inspection of the distribution of sources in Figure~\ref{fig:dist} that there 
are several clumps covering 10-20 arcsec in each sample; 
it is these that generate the high clustering amplitude.  
To make sure the signal was not
spurious, caused for example by difficulties in identifying single
galaxies from complex irregular galaxies, we visually inspected the
objects in each of these clumps.  The individual galaxies appeared
well separated and certainly not parts of single irregular objects. 
We conclude that the measurements represent real galaxy clustering.

\begin{figure*}
\vspace{8cm}  
\includegraphics{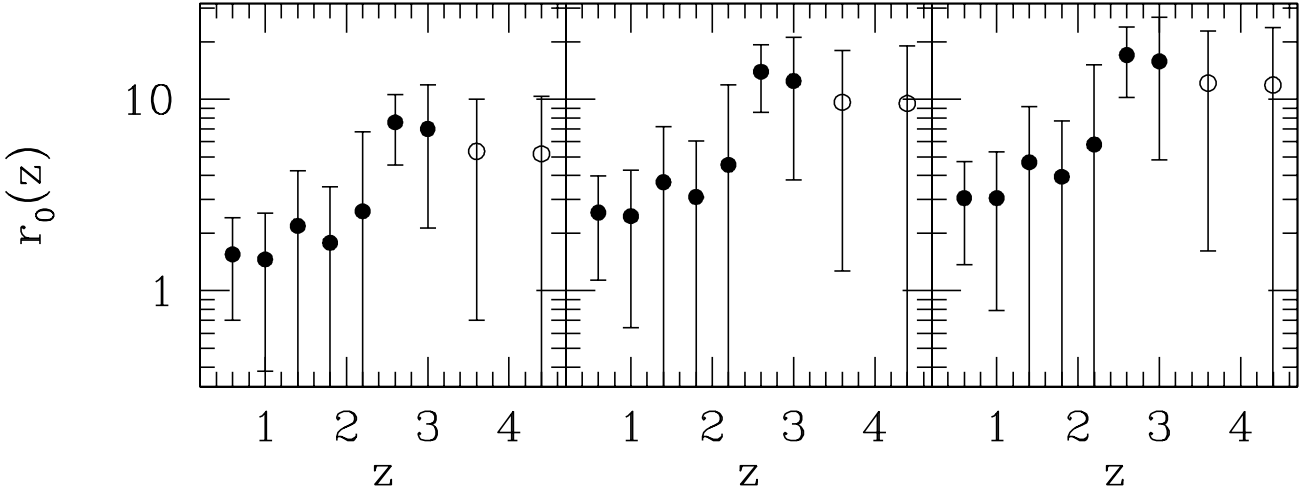} 
\caption{Trend of $r_0$ as a function of redshift as obtained
  from the analysis of the HDF photometric redshift catalogue by
  Fernandez-Soto et al., 1998. The left panel shows results for
  $\Omega_0=1$, $\Lambda=0$ and $h_0=1$, the panel in the middle shows 
results for $\Omega_0=0.4$, $\Lambda=0$ and $h_0=0.65$ while the right
  panel is for $\Omega_0=0.4$, $\Lambda=0.6$ and $h_0=0.65$. The filled
  points have been derived from considering redshift bins of width
  $\Delta z=0.4$, while the empty ones are for $\Delta z =0.8$.
\label{fig:fit} }
\end{figure*} 

If we assume a power-law form for $w(\theta)=A\theta^{1-\gamma}$,
we can estimate the parameters $A$ and $\gamma$, using a least-squares
fit to the data.  Given the large errors on $w$ we assumed a fixed
value of $\gamma=1.8$.
The small area of the HDF catalogue introduces a negative bias through
the integral constraint, $\int w^{est} d\Omega = 0 $. We allow for
this by fitting to $ A \theta^{1-\gamma} -C $, where $C = 25A$, 
  Furthermore, even though $w(\theta)$ has
been measured up to $\theta\simeq 0^{\circ}.05$, only angular scales
less than half the sample size are likely to be reliable (see also
Connolly et al., 1998), so we limit the fit to $\theta\simeq
0^{\circ}.02$. The dashed lines in Figure \ref{fig:w} represent the
best fit for each redshift interval; the best values for the amplitude
$A$ as a function of redshift are listed in Table~1.  These values coincide
within the errors with the results obtained by Connolly et al. in
their analysis of the clustering in the HDF for $z\simlt 1.4$.

\section{RELATION TO SPATIAL QUANTITIES}
The standard way of relating the angular two-point correlation
function $w(\theta)$ to the spatial two-point correlation function
$\xi(r,z)$ is by means of the relativistic Limber equation (Peebles,
1980):
\begin{eqnarray}
w(\theta)=2\:\frac{\int_0^{\infty}\int_0^{\infty}F^{-2}(x)x^4\Phi^2(x)
\xi(r,z)dx\:du}{\left[\int_0^{\infty}F^{-1}(x)x^2\Phi(x)dx\right]^2},
\label{eqn:limber} 
\end {eqnarray}
where $x$ is the comoving coordinate, $F(x)$ gives the correction for
curvature, and the selection function $\Phi(x)$ satisfies the relation
\begin{eqnarray}
{\cal N}=\int_0^{\infty}\Phi(x) F^{-1}(x)x^2 dx=\frac{1}{\Omega_s}\int_0^{\infty
}N(z)dz,
\label{eqn:Ndense} 
\end{eqnarray}
in which $\cal N$ is the mean surface density on a surface of solid angle
$\Omega_s$ and $N(z)$ is the number of objects in the given survey
within the shell ($z,z+dz$). 
Given the small range of angular scales sampled by the HDF we can
reasonably assume $\xi(r,z)$ to have the redshift dependent power-law 
form
\begin{eqnarray}
\xi(r,z)=\left(\frac{r}{r_0(z)}\right)^{-\gamma}
\label{eqn:power}
\end{eqnarray}
where all the dependence on $z$ is included in the correlation scale
length $r_0(z)$.
The physical separation between two sources separated by an angle
$\theta$ is given (in the small angle approximation) by:
\begin{eqnarray}
r\simeq\frac{1}{(1+z)}\;\left(\frac{u^2}{F^2}+x^2\theta^2\right)^{1/2}.
\label{eqn:r1}
\end{eqnarray} 

By including equations (\ref{eqn:Ndense}), (\ref{eqn:power}) and
(\ref{eqn:r1}) in equation (\ref{eqn:limber}) we then get the
following expression for the angular correlation function $w(\theta)$
\begin{eqnarray}
w(\theta)=\frac{H_{\gamma}\int_0^{\infty}N(z)^2P(\Omega_0,z)\;
(x(z)\;\theta)^{1-\gamma}r_0(z)^{\gamma}F(z)\; dz}
{\frac{c}{H_0}\;\left[\int_0^{\infty}N(z)\;dz\right]^2}
\label{eqn:w}
\end{eqnarray}
with
$H_{\gamma}=\Gamma[1/2]\Gamma[(\gamma-1)/2]/\Gamma[\gamma/2]=3.68$ in
the case of $\gamma=1.8$, $H_0$ the Hubble constant and $\Omega_0$ is
the density parameter, and $P = dx/dz$ .

If we consider a narrow redshift bin $\Delta z$ centred at some $\bar
z$ we can consider $N(z)$ constant in that interval. Under this
assumption the expression for the correlation length $r_0(\bar z)$ in
comoving coordinates is given by:
\begin{eqnarray}
r_0(\bar z)=\left(\frac{c\; A_{\Delta z}\;\Delta z}{H_0\;H_{\gamma}\;
x(\bar z)^{1-\gamma}P(\Omega_0,\bar z)\;F(\bar z)}\right)^{1/\gamma}
\label{eqn:r}
\end {eqnarray}
\noindent
where $A_{\Delta z}$ is the amplitude of the angular correlation
function $w(\theta)$ for a particular redshift interval.
Note that in practice the error in redshifts (true $\Delta z$) 
is the convolution of the measured $\Delta z$ 
with the error distribution. Assuming Gaussian error distribution,
this  will increase the effective $\Delta z$
in equation (8) by a factor $\sqrt{\frac{12 \sigma^2}{(\Delta z)^2}+1}$,
where $\Delta z$ is the bin width and $\sigma=0.1(z+1)$ is the $rms$
error on each redshift. 
\begin{table*}
\begin{center}
\begin{tabular}{ccccccc}
$\Delta z$& $\bar z $ & $N_{gal}$ & $A$ & $r_0$ (a) & $r_0$ (b) & $r_0$ (c)\\
\hline
$0.4-0.8$ & $0.6$ &$151$ & $(0.7\pm 0.4)\cdot 10^{-3}$ & $1.55\pm
0.85$ & $2.56\pm 1.42$ & $3.05\pm 1.68$\\
$0.8-1.2$ & $1.0$ &$173$ & $(0.6\pm 0.4)\cdot 10^{-3}$ & $1.46\pm
1.08$ & $2.45\pm 1.81$ & $3.05\pm 2.26$\\
$1.2-1.6$ & $1.4$ &$165$ & $(1.2\pm 1.0)\cdot 10^{-3}$ 
& $2.18\pm 2.06$ & $3.69\pm 3.52$ & $4.69\pm 4.47$\\
$1.6-2.0$ & $1.8$ &$161$ & $(0.8\pm 0.7)\cdot 10^{-3}$ & $1.78\pm
1.71$ & $3.09\pm 2.95$ & $3.94\pm 3.77$\\
$2.0-2.4$ & $2.2$ &$66$ & $(1.6\pm 2.3)\cdot 10^{-3}$ & $2.60\pm 4.18$ 
& $4.55\pm 7.31$ & $5.80\pm 9.34$\\
$2.4-2.8$ & $2.6$ &$36$ & $(11\pm 4.0)\cdot 10^{-3}$ & $7.58\pm
3.04$ & $13.9\pm 5.36$ & $17.0\pm 6.83$\\
$2.8-3.2$ & $3.0$ &$36$ & $(9.8\pm 7.0)\cdot 10^{-3}$ & $7.00\pm 4.87$ 
& $12.4\pm 8.66$ & $15.8\pm 10.9$\\
$3.2-4.0$ & $3.6$ &$37$ & $(5.6\pm 5.0)\cdot 10^{-3}$ & $5.35\pm 4.65$ 
& $9.65\pm 8.38$ & $12.2\pm 10.5$\\
$4.0-4.8$ & $4.4$ &$39$ & $(5.9\pm 5.0)\cdot 10^{-3}$ & $5.19\pm 5.17$ 
& $9.52\pm 9.49$ & $11.8\pm 11.8$\\
\end{tabular}
\end{center}
\caption{Results for each redshift range, $\Delta z$: mean redshift $\bar z$; 
number of galaxies; amplitude of the angular correlation function at $1^\circ$,
$A$; the spatial clustering amplitude $r_0$ for three different
cosmologies 
(a)~$\Omega_0=1,\;\Lambda=0,\;h_0=1$; 
(b)~$\Omega_0=0.4,\;\Lambda=0,\;h_0=0.65$; 
(c)~$\Omega_0=0.4,\;\Lambda=0.6,\;h_0=0.65$.}
\end{table*}

The geometry of space will determine the comoving coordinate
$x$, the curvature correction factor $F(x)$ and
the quantity $P(\Omega_0,z)$. In particular, for a Universe with
generic density 
parameter $\Omega_0$ and cosmological constant $\Lambda=0$ (see 
e.g. Magliocchetti et al., 1998; Treyer \&
Lahav, 1996) we have:
\begin{eqnarray}
x=\frac{2c}{H_0}\left[\frac{\Omega_0 z-(\Omega_0-2)(1-\sqrt{1+\Omega_0
      z})}{\Omega^2_0 (1+z)}\right],
\label{eqn:x1}
\end{eqnarray}
\begin{eqnarray}
F(x)=\left[1-\left(\frac{H_0 x}{c}\right)^2(\Omega_0 -1)\right]^{1/2}
\label{eqn:F1}
\end {eqnarray}
and
\begin{eqnarray}
P(\Omega_0,z)=\frac{\Omega_0^2(1+z)^2(1+\Omega_0 z)^{1/2}}{4(\Omega_0-1)[(1+\Omega_0
  z)^{1/2}-1]+\Omega_0^2(1-z)+2\Omega_0 z}
\label {eqn:P1}
\end{eqnarray}
\\
In the  case of a cosmological constant $\Lambda \ne 0$ with
$\Omega_0+\Lambda=1$ (flat space) we have $F(x)=1$,
\begin{eqnarray}
x=\frac{c}{H_0}\Omega_0^{-1/2}\int_0^z\frac{dz}{\left[(1+z)^3+
\Omega_0^{-1}-1\right]^{1/2}},
\label {eqn:x2}
\end{eqnarray}
(see Peebles, 1984; Magliocchetti et al., 1998; Treyer \& Lahav, 1996) and
\begin{eqnarray}
P(\Omega_0,z)=\Omega_0^{1/2}[(1+z)^3+\Omega_0^{-1}-1]^{1/2}.
\label {eqn:P2}
\end{eqnarray}

Using the values of
$A_{\Delta z}$ obtained in section 3, and once again
using $\gamma=1.8$ we find the values for the correlation length $r_0$
listed in Table 1, according to
the different cosmologies used in the deprojection analysis. 
Figure \ref{fig:fit} shows the trend of $r_0$ as a function of
the redshift $z$ for the three cosmological models $\Omega_0=1$,
$\Lambda=0$, $h_0=1$ (left panel), $\Omega_0=0.4$, $\Lambda=0$,
$h_0=0.65$ (central panel) and $\Omega_0=0.4$, $\Lambda=0.6$,
$h_0=0.65$ (right panel). Given the uncertainties in the determination of the
photometric redshifts at high $z$'s (see section 3), 
for $z < 3.2$ we have plotted the measurements coming from a
$\Delta z=0.4$ binning (filled circles), while for $z\simgt 3.2$ we
only considered broader bins of width $\Delta z=0.8$ (empty circles). 

The figure shows that $r_0$ is roughly constant with look-back time
for $z\simlt 2.4$. Above this redshift, the clustering amplitude
increases by more than a factor of 2.
Our measurements at high redshifts are in excellent agreement with the
values obtained in the analysis of the clustering of the Lyman Break
Galaxies (e.g., for $\Omega_0=1$, $r_0=4\pm 1$ - Adelberger
et al., 1998 and $r_0=2.1\pm 0.5$ - Giavalisco et al., 1998,
according to the particular sample used).
Given the small volume sampled by the HDF data, the uncertainties are
large, and the exact form of the evolution of $r_0$ is not well
determined. The measurements are consistent with a slow decline followed by
a smooth rise in amplitude above $z\simgt 2.4$, as predicted by some
galaxy formation models. 
From our measurements alone we cannot rule out a simple smooth
increase of $r_0$ with $z$, but compared to local surveys
our measurements  at $z\sim 1$ require a significant drop in 
the correlation length between $z=0$ and $z=1$. 

\section{CONCLUSIONS}

We have calculated the angular correlation function of galaxies in the
HDF image as a function of redshift up to $z\simeq 4.8$.  We use the
catalogue of photometric redshifts obtained by Fernandez-Soto et al.,
1998 to select redshift limited sub-samples of width $\Delta z=0.4$,
and $\Delta z=0.8$ between $z \simeq 0.4$ and $z\simeq 4.8$. 
The results show that, while for $z\simlt 2.4 $ the clustering
amplitude is roughly constant, for $z\simgt 2.4 $ there is a
significant increase in amplitude.

Converting the projected clustering amplitude into 
the correlation length $r_0$ at
different redshifts, we find that $r_0(z)$ is roughly constant as
a function of look-back time until $z\simeq 2.4$; at higher redshifts
the clustering amplitude rises to much higher values than for lower
z's. Our high-redshift
measurements are in good agreement with those obtained from the
analysis of the Lyman Break Galaxies (Adelberger et al., 1998;
Giavalisco et al., 1998). 

Under the assumption of linear evolution of mass fluctuations, we
would expect a slow decrease in $r_0$ towards higher redshifts.
However, we measure the clustering of galaxies, which are biased
tracers of the mass. The bias level is unlikely to be a constant as a
function of redshift; any galaxy seen beyond $z\sim2.4$ has formed stars
at an epoch earlier than most galaxies, and so is likely to be biased
relative to an ``average'' galaxy.
Also, the HDF galaxy sample is selected on observed frame $I$ band and
so different populations of galaxies are selected in the different
redshift ranges: at $z\sim 1$ the selection is roughly rest-frame B
band, but at higher redshift samples are selected on
rest-frame UV flux, which will preferentially select the galaxies with
higher star-formation rates.  

A further complication is caused by the effects of gravitational
lensing. We expect that structure in the foreground mass distribution
will introduce an extra component of clustering through the
gravitational lensing magnification bias (Villumsen et al 1997).
The amplitude of this effect depends on the 
amplitude of mass fluctuations. 

Hence the interpretation of galaxy clustering in the HDF at high
redshifts is not at all straightforward.  Theoretical models (see
  for instance Kauffmann et al., 1998; Baugh et al., 1998) predict a
  trend for the evolution of the clustering signal with look-back time
  that is similar to what detected in our analysis, but
  their results are strongly model-dependent. We will present a detailed
analysis and comparison to models in a future paper.

\vspace{0.3cm}
\noindent
{\bf ACKNOWLEDGEMENTS}
MM acknowledges support from the Isaac Newton Scholarship. Jarle
Brinchmann is warmly thanked for very helpful and
stimulating discussions.

\end{document}